\documentclass[12pt]{article}
\usepackage{amssymb}
\def\a{\alpha}

\def\g{\gamma}

\def\l{\lambda}

\def\s{\sigma}

\def\up{\uparrow}
\def\down{\downarrow}

\makeatletter
\renewcommand\thesection{\@arabic\c@section}
\renewcommand\thesubsection{\thesection.\@arabic\c@subsection}
\makeatother

\newcommand{\sect}[1]{\setcounter{equation}{0}\section{#1}}

\begin{document}

%************************** Text Begins here ******************************

%  Greek letters

\def\R{\overline{R}}
% Shorthands for \begin{equation} and the like

%\newcommand{\sect}[1]{\setcounter{equation}{0}\section{#1}}
%\renewcommand{\theequation}{\thesection.\arabic{equation}}

\baselineskip=20pt

%%%%%%%%%%%%%%%%%%%%%%%%%%%%%%%%%%%%%%%%%%%%%%%%%%%%%%%%%%%%
%                                                          %
%  Title page                                              %
%                                                          %
%%%%%%%%%%%%%%%%%%%%%%%%%%%%%%%%%%%%%%%%%%%%%%%%%%%%%%%%%%%%
\newfont{\elevenmib}{cmmib10 scaled\magstep1}
\newcommand{\preprint}{
   %\begin{flushleft}
    % \elevenmib Yukawa\, Institute\, Kyoto\\
   %\end{flushleft}\vspace{-1.3cm}
   \begin{flushright}\normalsize
%     {\tt hep-th/0701042}\\ 2006
   \end{flushright}}
\newcommand{\Title}[1]{{\baselineskip=26pt
   \begin{center} \Large \bf #1 \\ \ \\ \end{center}}}
\newcommand{\Author}{\begin{center}
   \large \bf
 Shao-You Zhao${\,}^a$  and Yao-Zhong Zhang${\,}^{a,b}$
 \end{center}}
\newcommand{\Address}{\begin{center}
${}^a$  Department of Mathematics, University of Queensland,
            Brisbane, QLD 4072, Australia\\
${}^b$ Physikalisches Institut, Universit\"at Bonn, D-53115 Bonn, Germany

 E-mail: syz@maths.uq.edu.au, yzz@maths.uq.edu.au

 \end{center}}
   \newcommand{\Accepted}[1]{\begin{center}
   {\large \sf #1}\\ \vspace{1mm}{\small \sf Accepted for Publication}
   \end{center}}
\preprint \thispagestyle{empty}
\bigskip\bigskip\bigskip

\Title{ Supersymmetric Vertex Models with Domain Wall Boundary Conditions }
\Author

\Address \vspace{1cm}

\begin{abstract}
By means of the Drinfeld twists, we derive the determinant
representations of the partition functions for the $gl(1|1)$ and $gl(2|1)$
supersymmetric vertex models with domain wall boundary conditions. In the
homogenous limit, these determinants degenerate to simple functions.

\end{abstract}

\vspace{1truecm}

%\noindent {\it PACS:} 03.65.Fd; 04.20.Jb; 05.30.-d; 75.10.Jm


\newpage
\sect{Introduction}

The six vertex model on a finite square lattice with the so-called domain
wall (DW) boundary condition was first proposed by Korepin in
\cite{Korepin82}, where recursion relations of the partition functions of the
were derived.  In \cite{Izergin87,Izergin92} it was found that the partition
 functions of the DW model can be represented as
determinants. Taking the homogenous limit of the spectral
parameters, Sogo found that the partition function satisfies the
Toda differential equations \cite{Sogo93}. Then by using the
equations Korepin and Zinn-Justin obtained the bulk free energy of
the system \cite{Korepin00}. The determinant representations of
DW partition functions are uesful in solving some
pure mathematical problems, such as the problem of
alternating sign matrices \cite{Kuperburg96}. By using the fusion
method, Caradoc, Foda and Kitanine obtained the determinant
expressions of the partition functions for the spin-$k/2$ vertex
models with the DW boundary conditions \cite{Caradoc}. In
\cite{Bleher05}, Bleher and Fokin obtained the large $N$
asymptotics of the DW six-vertex model in the disordered phase.

Supersymmetric integrable models based on superalgebras
are important for describing strongly correlated electronic systems of
high $T_c$ superconductivity
\cite{Kul82,Korepin92,Essler94,Bra95,Frahm-forU,Martins-forU,Katrina-forU}.
%The eigenvalue problems of several supersymmetric fermion models were solved
%exactly by means of algebraic Bethe ansatz
%\cite{Korepin92,Essler94,Frahm-forU,Martins-forU,Katrina-forU}.
In this paper, we investigate the $gl(1|1)$ and $gl(2|1)$
supersymmetric vertex models on a $N\times N$ square lattice with
DW boundary conditions. By using the approach of Drinfeld twists
\cite{Maillet96,zsy04,zsy05}, we derive the determinant
representations of the DW partition functions for the two systems.
We find that the partition functions degenerate to simple
functions in the homogenous limit.

The framework of the paper is as follows. In section 2, we
obtain the determinant representation of the partition function of the
$gl(1|1)$ vertex model with DW boundary condition. Then we
derive the homogenous limit of the partition function.
In section 3, we solve the $gl(2|1)$ supersymmetric vertex
model with the DW boundary conditions, by deriving its DW partition
functions exactly. In section 4, we present some discussions.

\sect{$gl(1|1)$ vertex model with DW boundary condition}

In this section, we study the DW boundary condition for
the $gl(1|1)$ vertex model on a $N\times N$ square lattice.

\subsection{Description of the model}

Let $V$ be the 2-dimensional irreducible $gl(1|1)$-module and
$R\in End(V\otimes V)$ the $R$-matrix associated with this
representation. In this paper, we choose the FB grading for $V$,
i.e. $[1]=1,[2]=0$. The the $R$-matrix satisfies the graded
Yang-Baxter equation (GYBE)
\begin{eqnarray}
R_{12}R_{13}R_{23}= R_{23}R_{13}R_{12}, \label{eq:YBE}
\end{eqnarray}
where $R_{ij}\equiv R_{ij}(\lambda_i,\lambda_j)$ with spectral
parameters specified by $\lambda_i$. Explicitly,
\begin{eqnarray}
 R_{12}(\lambda_1,\lambda_2)%\nonumber\\
% &=& \nonumber\\
          &=&\left(\begin{array}{ccccccccc}
 c_{12}&0&0 & 0\\
 0&a_{12}& b_{12}&0 \\
 0&b_{12}&a_{12}&0\\
 0&0&0&1
 \end{array} \right), \label{de:R-gl11}
\end{eqnarray}
where
\begin{eqnarray}
&& a_{12}=a(\lambda_1,\lambda_2)\equiv
  {\lambda_1-\lambda_2\over \lambda_1-\lambda_2+\eta},\quad \quad
b_{12}=b(\lambda_1,\lambda_2)\equiv
  {\eta\over \lambda_1-\lambda_2+\eta},\quad\quad
\nonumber\\
&&c_{12}=c(\lambda_1,\lambda_2)\equiv
{\lambda_1-\lambda_2-\eta\over \lambda_1-\lambda_2+\eta}
\end{eqnarray}
and $\eta$ is the crossing parameter which could be normalized to 1 in
the rational cases we are considering.

%In terms of the matrix elements defined by
%\begin{equation}
%R(\l)(v^{i'}\otimes
%v^{j'})=\sum_{i,j}R(\l)^{i'j'}_{ij}(v^{i}\otimes v^{j}),
%\end{equation}
%the GYBE reads
%\begin{eqnarray}
%&&
%\sum_{i',j',k'}R(\l_1-\l_2)^{i'j'}_{ij}R(\l_1-\l_3)^{i''k'}_{i'k}
%R(\l_2-\l_3)^{j''k''}_{j'k'}
%    (-1)^{[j']([i']+[i''])}\nonumber\\
%&=&\sum_{i',j',k'}R(\l_2-\l_3)^{j'k'}_{jk}R(\l_1-\l_3)^{i'k''}_{ik'}
%R(\l_1-\l_2)^{i''j''}_{i'j'}
%    (-1)^{[j']([i]+[i'])}.
%\end{eqnarray}

In the following, we study the $gl(1|1)$ 6-vertex model on a 2-d
$N\times N$ square lattice. At any site of the lattice, there may
be a vacuum $\phi$ or a single fermion $\rho$.
A vertex configuration in the lattice is constructed
by two nearest particle states in a horizontal line and two
nearest particle states in a  vertical line. For the present
model, there are altogether 6 possible weights corresponding to a
vertex configuration. The possible configurations and their
corresponding Boltzmann weights $w_i$ are given as follows:
\newpage
$$
\begin{picture}(50,50)(0,0)
\put(-135,5){\line(-1,0){30}} \put(-150,20){\line(0,-1){30}}
\put(-154,-23){$\phi$}\put(-154,23){$\phi$}
\put(-175,0){$\phi$}\put(-133,0){$\phi$} \put(-155,-45){$w_1$}
\put(-35,5){\line(-1,0){30}} \put(-50,20){\line(0,-1){30}}
\put(-54,-23){$\rho$}\put(-54,23){$\rho$}
\put(-75,0){$\rho$}\put(-33,0){$\rho$} \put(-55,-45){$w_2$}
\put(65,5){\line(-1,0){30}} \put(50,20){\line(0,-1){30}}
\put(44,-23){$\phi$}\put(44,23){$\phi$}
\put(25,0){$\rho$}\put(67,0){$\rho$} \put(45,-45){$w_3$}
\put(165,5){\line(-1,0){30}} \put(150,20){\line(0,-1){30}}
\put(144,-23){$\rho$}\put(144,23){$\rho$}
\put(125,0){$\phi$}\put(167,0){$\phi$} \put(145,-45){$w_4$}
\end{picture}
$$\\[3mm]
%%%%%%%%%%%%%%%%%%%%%%%%%%%%%%%%%%%%%%%%%%%%%%%%%%%%
$$
\begin{picture}(50,50)(0,0)
\put(-135,5){\line(-1,0){30}} \put(-150,20){\line(0,-1){30}}
\put(-154,-23){$\phi$}\put(-154,23){$\rho$}
\put(-175,0){$\rho$}\put(-133,0){$\phi$} \put(-155,-45){$w_5$}
\put(-35,5){\line(-1,0){30}} \put(-50,20){\line(0,-1){30}}
\put(-54,-23){$\rho$}\put(-54,23){$\phi$}
\put(-75,0){$\phi$}\put(-33,0){$\rho$} \put(-55,-45){$w_6$}
\end{picture}
$$\\[2mm]
$$
\begin{picture}(50,50)(0,0)
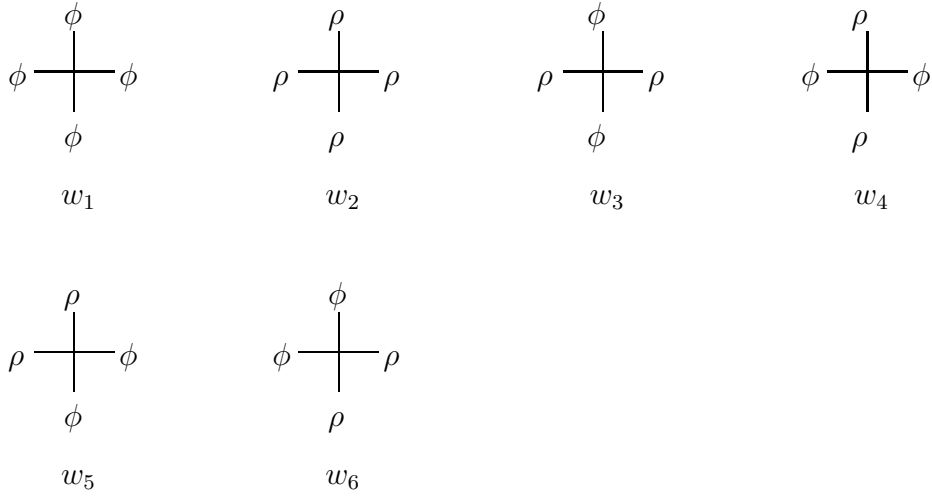
\put(-175,30){\small{\bf Figure 1.} Vertex configurations and
their Boltzmann weights for the $gl(1|1)$ vertex model.}
\end{picture}
$$
\vskip3mm
$$
\begin{picture}(50,50)(0,0)
\put(20,30){\line(-1,0){100}} \put(20,10){\line(-1,0){100}}
\put(20,-10){\line(-1,0){100}} \put(20,-30){\line(-1,0){100}}
 \put(0,50){\line(0,-1){100}}\put(-20,50){\line(0,-1){100}}
 \put(-40,50){\line(0,-1){100}}\put(-60,50){\line(0,-1){100}}
 \put(-95,27){$\phi$} \put(-95,7){$\phi$}
 \put(-95,-13){$\phi$} \put(-95,-33){$\phi$}
 \put(27,27){$\rho$}\put(27,7){$\rho$}
 \put(27,-13){$\rho$}\put(27,-33){$\rho$}
 \put(-5,57){$\phi$}\put(-25,57){$\phi$}
 \put(-45,57){$\phi$}\put(-65,57){$\phi$}
 \put(-5,-60){$\rho$} \put(-25,-60){$\rho$}
 \put(-45,-60){$\rho$} \put(-65,-60){$\rho$}
\put(-70,42){{\small$z_1$}}\put(-50,42){{\small$z_2$}}
\put(-35,42){{\small$\ldots$}}\put(-12,42){{\small$z_N$}}
\put(-77,33){{\small$\l_1$}}\put(-77,13){{\small$\l_2$}}
\put(-74,-5){{\small$\vdots$}}\put(-77,-27){{\small$\l_N$}}
\end{picture}
$$ \vskip17mm
$$
\begin{picture}(50,50)(0,0)
\put(-175,30){\small{\bf Figure 2.}
DW boundary condition for the $gl(1|1)$ vertex model.}
\end{picture}
$$
One may check that these weights preserve the fermion numbers. If
we assign horizontal lines parameters $\{\lambda_j\}$ and vertical
lines parameters $\{z_k\}$, and let
\begin{eqnarray}
w_1=1,\quad w_2=c(\lambda,z),\quad w_3=w_4=a(\lambda,z),\quad
w_5=w_6=b(\lambda,z), \label{de:w-gl11}
\end{eqnarray}
 then the Boltzmann weights correspond to the elements of the $gl(1|1)$
$R$-matrix (\ref{de:R-gl11}).
%Thus, the vertex configurations of any horizontal or vertical line
%form a monodromy matrix (\ref{de:T-gl11}) and then the Hamiltonian
%of the corresponding 1-d lattice model will be the free fermion model.

We now propose the $gl(1|1)$ vertex model with DW boundary condition.
Similar to \cite{Korepin82,Izergin92}, if the ends of the square lattice
satisfy the special boundary condition, that is states of all left
and top ends are vacuum $\phi$ while states of all right and bottom ends
are single fermion $\rho$, we then call the
boundary condition the DW boundary condition.

A configuration with the DW boundary condition is shown in figure
2. In this figure, $\lambda_i$ and $z_j$ are parameters associated
with the horizontal and vertical lines, respectively. The
partition function for this system is then given by
\begin{eqnarray}
Z_N=\sum\prod_{i=1}^6 w_i^{n_i},
\end{eqnarray}
where the summation is over all possible configurations, $n_i$ is
the number of configurations with the Boltzmann weight $w_i$. By
means of the R-matrix, the partition function may be rewritten as
\begin{eqnarray}
Z_N=\prod_{i=0}^{N-1}\rho_{(N-i)}\prod_{j=0}^{N-1}\phi_{(N-j)}
\prod_{k=1}^N\prod_{l=1}^N R_{kl}(\lambda_k,z_l)
 \prod_{j=1}^{N}\rho_{(j)}\prod_{i=1}^{N} \phi_{(i)}.
 \label{eq:Z-gl11}
\end{eqnarray}
Here $\rho_{(i)}$ ($\phi_{(i)}$) stands for the state $\rho$ ($\phi$)
in $i$th horizontal (vertical) line.

Following Korepin, we use the transfer matrix to arrange the
partition function. Define the monodromy matrix $T_k(\lambda_k)$
along the horizontal lines
\begin{eqnarray}
T_k(\lambda_k)=R_{k,N}(\lambda_k,z_N)R_{k,N-1}(\lambda_k,z_{N-1})\ldots
R_{k,1}(\lambda_k,z_1)
 \equiv\left(\begin{array}{cc}A(\lambda_k)& B(\lambda_k)\\ C(\lambda_k)&
 D(\lambda_k)\end{array}\right)_{(k)}. \label{de:T-gl11}
\end{eqnarray}
therefore, in terms of $T_k(\lambda_k)$ the partition function is
equal to
\begin{eqnarray}
Z_N &=&\prod_{i=0}^{N-1}\langle 1|_{(N-i)}\prod_{j=0}^{N-1}\langle
0|_{(N-j)} \prod_{k=1}^N T_k(\lambda_k)
 \prod_{j=1}^{N}|1\rangle_{(j)}\prod_{i=1}^{N}
 |0\rangle_{(i)}\nonumber\\
 &=&\prod_{i=0}^{N-1}\langle 1|_{(N-i)}C(\lambda_1)C(\lambda_2)\ldots
 C(\lambda_N)\prod_{i=1}^{N} |0\rangle_{(i)}. \label{eq:Z-gl111}
\end{eqnarray}
Here we have introduced the notation
\begin{eqnarray}
|0\rangle =\left(\begin{array}{c} 0\\1\end{array}\right),
 \quad\quad
|1\rangle =\left(\begin{array}{c} 1\\0\end{array}\right)
\label{de:states-gl11}
\end{eqnarray}
to denote the particle states $\phi, \rho$, respectively.

\subsection{Exact solution of the partition function}

In this subsection, we will compute the partition function
(\ref{eq:Z-gl111}) for the $gl(1|1)$ vertex model with DW boundary
conditions by using the Drinfeld twist approach.

Let us remark that even though we believe that Drinfeld twists do
exist for all algebras, so far only those related to $A$-type
(super) algebras and to $XYZ$ model have been explictly
constructed \cite{Albert00,zsy04,zsy05,zsy0502,Albert0005}.

Let $\sigma$ be any element of the permutation group $S_N$. We
then define the following lower-triangular matrix \cite{zsy05,
zsy0502}
\begin{eqnarray}
F_{1,\ldots N}=\sum_{\sigma\in {\cal S}_N}
   \sum_{\alpha_{\sigma(1)}\ldots\alpha_{\sigma(N)}}^{\quad\quad *}
   \prod_{j=1}^N P_{(\sigma(j))}^{\alpha_{\sigma(j)}}
   S(c,\sigma,\alpha_\sigma)R_{1\ldots N}^\sigma, \label{de:F}
\end{eqnarray}
where $P^\a_{(k)}$ has the elements
$(P^\alpha_{(k)})_{mn}=\delta_{\alpha m}\delta_{\alpha n}$ at the
$k$th space with root indices $\a=1,2$, the sum $\sum^*$ is taken
over all non-decreasing sequences of the labels
$\alpha_{\sigma(i)}$:
\begin{eqnarray}
&& \alpha_{\sigma(i+1)}\geq \alpha_{\sigma(i)},
 \quad\mbox{if}\quad
              \sigma(i+1)>\sigma(i), \nonumber\\
&& \alpha_{\sigma(i+1)}> \alpha_{\sigma(i)},
 \quad \mbox{if}\quad
              \sigma(i+1)<\sigma(i) \label{cond:F}
\end{eqnarray}
and the c-number function $S(c,\sigma,\alpha_\sigma)$ is given by
\begin{eqnarray}
S(c,\sigma,\alpha_\sigma)\equiv \exp\left\{{1\over2}
 \sum_{l>k=1}^N\left(1-(-1)^{[\alpha_{\sigma(k)}]}\right)
 \delta_{\alpha_{\sigma(k)},\alpha_{\sigma(l)}}
    \ln(1+c_{\sigma(k)\sigma(l)})\right\}. \label{de:S}
\end{eqnarray}
In (\ref{de:F}),  $R^\sigma_{12\ldots N}$ is the
$N$-fold $R$-matrix which can be decomposed in terms of elementary
$R$-matrices (\ref{de:R-gl11}) by the decomposition law
\begin{eqnarray}
 R^{\sigma'\sigma}_{1\ldots N}
 =R^{\sigma}_{\sigma'(1\ldots N)}
  R^{\sigma'}_{1\ldots N}.\label{eq:R-RR}
\end{eqnarray}
We showed in \cite{zsy04} that the $F$-matrix is nondegenerate and
satisfies the relation
\begin{eqnarray}
F_{\sigma(1\ldots N)}(z_{\sigma(1)},\ldots,z_{\sigma(N)})
  R_{1\ldots N}^\sigma(z_1,\ldots,z_N)
 =F_{1\ldots N}(z_1,\ldots,z_N). \label{eq:R-F-N}
\end{eqnarray}
The nondegeneracy of the $F$-matrix  means that its column vectors
form a complete basis, which is called the $F$-basis. The
nondegeneracy also ensures that the $F$-matrix is invertible. The
inverse is given by \cite{zsy05}
\begin{equation}
F^{-1}_{1\ldots N}=F^*_{1\ldots N}\prod_{i<j}\Delta_{ij}^{-1}
\label{eq:F-inverse}
\end{equation}
with
\begin{eqnarray}
\Delta_{ij}=\mbox{diag}\left((1+c_{ij})(1+c_{ji}),a_{ji},a_{ij},1\right)
\end{eqnarray}
and
\begin{eqnarray}
F^*_{1\ldots N}&=&\sum_{\sigma\in {\cal S}_N}
   \sum_{\alpha_{\sigma(1)}\ldots\alpha_{\sigma(N)}}^{\quad\quad **}
   S(c,\sigma,\alpha_\sigma)R_{\sigma(1\ldots N)}^{\sigma^{-1}}
   \prod_{j=1}^N P_{(\sigma(j))}^{\alpha_{\sigma(j)}}.
    \label{de:F*}  \nonumber\\
\end{eqnarray}
Here the sum $\sum^{**}$ is taken over all possible $\alpha_i$
which satisfies the following non-increasing constraints:
\begin{eqnarray}
&& \alpha_{\sigma(i+1)}\leq \alpha_{\sigma(i)},
 \quad\mbox{if}\quad
              \sigma(i+1)<\sigma(i), \nonumber\\
&& \alpha_{\sigma(i+1)}< \alpha_{\sigma(i)},\quad \mbox{if}\quad
              \sigma(i+1)>\sigma(i). \label{cond:F*}
\end{eqnarray}

Working in the $F$-basis, the entries of the monodromy matrix
(\ref{de:T-gl11}) can be simplified to symmetric forms, e.g. the
lower entry $C(\lambda)$ becomes \cite{zsy05}
\begin{eqnarray}
\tilde C(\lambda)&\equiv& F_{12\ldots N}
C(\lambda)F^{-1}_{12\ldots N}
 \nonumber\\
 &=&\sum_{i=1}^Nb(\lambda,z_i)E^{12}_{(i)}\otimes_{j\ne i}
 \mbox{diag}\left(2a(\lambda,z_j),1\right)_{(j)}, \label{eq:C-gl11}
\end{eqnarray}
where $E^{\alpha \beta}_{(l)}$ $(\alpha,\beta=1,2; l=1,2,\ldots,
N)$ are generators of the superalgebra $gl(1|1)$ at the site $l$.
From (\ref{eq:C-gl11}), one can see that in the F-basis all
compensating terms (polarization clouds) in the original
expression of $C(\lambda)$ in terms of local generators disappear
from $\tilde C(\lambda)$.

Applying the $F$-matrix and its inverse to the states
$|0\rangle_{(1)}\otimes\cdots\otimes |0\rangle_{(N)}$ and
$\langle1|_{(N)}\otimes\cdots\otimes \langle1|_{(1)}$, we have
\begin{eqnarray}
&& F_{1\ldots N}|0\rangle_{(1)}\otimes\cdots\otimes
|0\rangle_{(N)}=|0\rangle_{(1)}\otimes\cdots\otimes |0\rangle_{(N)}\nonumber\\
&&\langle1|_{(N)}\otimes\cdots\otimes \langle1|_{(1)}F^{-1}_{1\ldots N}
 =\langle1|_{(N)}\otimes\cdots\otimes
\langle1|_{(1)}\prod_{i<j}(2a(z_i,z_j))^{-1}. \label{eq:dual-gl11}
\end{eqnarray}

Substituting (\ref{eq:C-gl11}) and (\ref{eq:dual-gl11}) into the
partition function (\ref{eq:Z-gl111}), we obtain
\begin{eqnarray}
Z_N&=&\prod_{i=0}^{N-1}\langle
1|_{(N-i)}C(\lambda_1)C(\lambda_2)\ldots
C(\lambda_N)\prod_{i=1}^{N}
 |0\rangle_{(i)}\nonumber\\
&=&\prod_{i=0}^{N-1}\langle 1|_{(N-i)}
 F^{-1}_{1\ldots N}F_{1\ldots N} C(\lambda_1)\ldots C(\lambda_N)
 F^{-1}_{1\ldots N}F_{1\ldots N} \prod_{i=1}^{N}|0\rangle_{(i)}\nonumber\\
&=&\prod_{i<j}(2a(z_i,z_j))^{-1}
 \prod_{i=0}^{N-1}\langle 1|_{(N-i)} \tilde C(\lambda_1)\ldots \tilde C(\lambda_N)
 \prod_{i=1}^{N}|0\rangle_{(i)}\nonumber\\
&=&\prod_{i<j}(2a(z_i,z_j))^{-1}
 \prod_{i=0}^{N-1}\langle1|_{(N-i)}\nonumber\\
& & \times 2^{{N(N-1)\over 2}}\sum_{i_1<\ldots< i_{N}}
B_{N}(\lambda_1,\ldots,\lambda_{N}|z_{i_1},\ldots,z_{i_{N}})
 E_{(i_1)}^{12}\ldots
E_{(i_{n})}^{12}\,\prod_{i=1}^{N}|0\rangle_{(i)} \nonumber\\
&=&\prod_{i<j}a^{-1}(z_i,z_j)B_{N}(\lambda_1,\ldots,\lambda_{N}|z_{1},\ldots,z_{N}),
\end{eqnarray}
where
\begin{eqnarray}
B_{N}(\lambda_1,\ldots,\lambda_{N}|z_{1},\ldots,z_{N})
 &=&
 \sum_{\sigma\in {\cal S}_N}\mbox{sign}(\sigma)
  \prod_{k=1}^N b(\lambda_k,z_{\sigma(k)})
  \prod_{l=k+1}^N a(\lambda_k,z_{\sigma(l)})\nonumber\\
  &=&\mbox{det}{\cal B}(\{\underline{\lambda}\},\{\underline{z}\})% \quad
 %(k=1,\dots n,\, j=i_1,\ldots,i_n)
  \label{eq:B_n}
\end{eqnarray}
with ${\cal B}(\{\underline{\lambda}\},\{\underline{z}\})$ being a
$N\times N$ matrix with elements
\begin{equation}
\left({\cal
B}(\{\underline{\lambda}\},\{\underline{z}\})\right)_{\alpha\beta}
 =b(\lambda_\alpha,z_\beta)
  \prod_{\gamma=1}^{\alpha-1}a(\lambda_\gamma,z_\beta).
\end{equation}
Therefore the partition function of the $gl(1|1)$ vertex model is
given by the following determinant of the $N\times N$ matrix,
\begin{eqnarray}
Z_N=\prod_{i<j}a^{-1}(z_i,z_j)\mbox{det}{\cal
B}(\{\underline{\lambda}\},\{\underline{z}\}).
\label{eq:Z-gl11-deter}
\end{eqnarray}

\subsection{Homogenous limit of the partition function}
In this subsection, we discuss the homogenous limit, i.e. when
$\lambda_1=\lambda_2=\ldots=\lambda_N$ and $ z_1=z_2=\ldots=z_N$,
of the partition function (\ref{eq:Z-gl11-deter}).

For later convenience, we rewrite the inhomogeneous partition
function more explicitly as
\begin{eqnarray}
&&Z_N=\prod_{i<j}{z_i-z_j+\eta\over z_i-z_j}\nonumber\\
 &&\times\left|\begin{array}{ccccc}
 b(\lambda_1,z_1) &\cdots&
 b(\lambda_1,z_k) &\cdots&
 b(\lambda_1,z_N)\\[2mm]
  \displaystyle b(\lambda_2,z_1)a(\lambda_1,z_1) &\cdots&
 \displaystyle b(\lambda_2,z_k)a(\lambda_1,z_k) &\cdots&
 \displaystyle b(\lambda_2,z_N)a(\lambda_1,z_N)\\
\vdots &\vdots
  &\vdots&\vdots&\vdots\\
 \displaystyle b(\lambda_N,z_1)\prod_{\g=1}^{N-1} a(\lambda_\g,z_1) &\cdots&
 \displaystyle b(\lambda_N,z_k)\prod_{\g=1}^{N-1}a(\lambda_\g,z_k) &\cdots&
 \displaystyle b(\lambda_N,z_N)\prod_{\g=1}^{N-1}a(\lambda_\g,z_N)\\
\end{array}
\right|\nonumber\\
\end{eqnarray}
It is easy to check that in the limit $\lambda_1 \rightarrow
\lambda, \lambda_2\rightarrow \lambda, \ldots,
\lambda_N\rightarrow \lambda$, we have
\begin{eqnarray}
&&Z_N=\prod_{i<j}{z_i-z_j+\eta\over z_i-z_j}
 \prod_{i=1}^N{\eta\over \lambda-z_i+\eta}
 \nonumber\\
 &&\times\left|\begin{array}{ccccc}
 1 &\cdots& 1 &\cdots& 1\\[2mm]
  \displaystyle a(\lambda,z_1) &\cdots&
 \displaystyle a(\lambda,z_k) &\cdots&
 \displaystyle  a(\lambda,z_N)\\
\vdots &\vdots
  &\vdots&\vdots&\vdots\\
 \displaystyle a ^{N-1}(\lambda,z_1) &\cdots&
 \displaystyle a^{N-1}(\lambda,z_k) &\cdots&
 \displaystyle a^{N-1}(\lambda,z_N)\\
\end{array}
\right|,\nonumber\\ \label{eq:Z-gl11-deter2}
\end{eqnarray}
where and below $a^k(\lambda,z_i)$ stands for the $k$-th power of
$a(\lambda,z_i)$. For the homogenous limit of the parameters
$z_k$, we first compute the limit $z_2\rightarrow z_1\equiv z$.
Taylor expanding the second column as $z_2\rightarrow z$,
$$a^k(\lambda,z_2)=a^k(\lambda,z)+(a^k(\lambda,z))'(z_2-z)+{\cal O}((z_2-z)^2),$$
where and below, $X'$, $X''$ and $X^{(n)}$ stand for the 1st, 2nd and $n$-th
order derivatives of $X$ with respect to the
parameter $z$, respectively, and subtracting the first column from the second,
(\ref{eq:Z-gl11-deter2}) becomes
\begin{eqnarray}
&&Z_N=(-\eta)\prod_{j=3}^N\left({z-z_j+\eta\over z-z_j}\right)^2
\prod_{j>i=3}{z_i-z_j+\eta\over z_i-z_j}
 \left({\eta\over \lambda-z+\eta}\right)^2\prod_{i=3}^N{\eta\over \lambda-z_i+\eta}
 \nonumber\\
 &&\times\left|\begin{array}{cccccccc}
 1 &0 &1&\cdots& 1\\[2mm]
 a(\lambda,z)& (a(\lambda,z))'& a(\lambda,z_3) &\cdots&
 a(\lambda,z_N)\\
\vdots &\vdots
  &\vdots&\vdots&\vdots\\
 a ^{N-1}(\lambda,z)& \left(a^{N-1}(\lambda,z)\right)' &
 a^{N-1}(\lambda,z_3) &\cdots&
 \displaystyle a^{N-1}(\lambda,z_N)\\
\end{array}
\right|. \label{eq:Z-gl11-deter3}
\end{eqnarray}
Then Taylor expanding the third column as $z_3\rightarrow z$,
$$
a^k(\lambda,z_3)=a^k(\lambda,z)+\left(a^k(\lambda,z)\right)'(z_3-z)
 +{1\over 2}\left(a^k(\lambda,z)\right)''(z_3-z)^2
 +{\cal O}((z_3-z)^3),
$$
and subtracting multiples of previous columns, we obtain
\begin{eqnarray}
&&Z_N={(-\eta)^3\over 2}\prod_{j=4}^N\left({z-z_j+\eta\over
z-z_j}\right)^3 \prod_{j>i=4}{z_i-z_j+\eta\over z_i-z_j}
 \left({\eta\over \lambda-z+\eta}\right)^3\prod_{i=4}^N{\eta\over \lambda-z_i+\eta}
 \nonumber\\
 &&\times\left|\begin{array}{cccccccc}
 1 &0 & 0&1&\cdots& 1\\[2mm]
 a(\lambda,z)& (a(\lambda,z))'&(a(\lambda,z))''& a(\lambda,z_4) &\cdots&
 a(\lambda,z_N)\\
\vdots &\vdots
  &\vdots&\vdots&\vdots\\
 a ^{N-1}(\lambda,z)& \left(a^{N-1}(\lambda,z)\right)' &
 \left(a^{N-1}(\lambda,z)\right)''&
 a^{N-1}(\lambda,z_4) &\cdots&
 \displaystyle a^{N-1}(\lambda,z_N)\\
\end{array}
\right|.\nonumber\\ \label{eq:Z-gl11-deter4}
\end{eqnarray}
Continuing with such process, we obtain, instead of the determinant
of the $N\times N$ matrix,
the following determinant of the $(N-1)\times (N-1)$ matrix for the partition
function:
\begin{eqnarray}
&&Z_N={(-\eta)^{\sum_{i=1}^{N-1}i}b^N(\lambda,z)\over
 \prod_{i=2}^{N-1} i!}
 \nonumber\\
 &&\times\left|\begin{array}{cccccccc}
(a(\lambda,z))'&(a(\lambda,z))'' &\cdots&
 (a(\lambda,z))^{(N-1)}\\
 \vdots
  &\vdots&\vdots&\vdots\\
  \left(a^{N-1}(\lambda,z)\right)' &
 \left(a^{N-1}(\lambda,z)\right)''&\cdots&
 \left( a^{N-1}(\lambda,z_N)\right)^{(N-1)}\\
\end{array}
\right|.\nonumber\\ \label{eq:Z-gl11-deter5}
\end{eqnarray}
To simplify the determinant in (\ref{eq:Z-gl11-deter5}) further,
we investigate its elements. Computing
$\left(a^{j}(\lambda,z)\right)^{(n)}$, we obtain
\begin{eqnarray}
\left(a^{j}(\lambda,z)\right)^{(n)}=\sum_{k=1}^n f_k
\prod_{l=0}^{k-1}
 (j-l)a^{j-k}(\lambda,z), \label{eq:a-derivation}
\end{eqnarray}
where $f_k$ are functions of $\{(a(\lambda,z))^{(l)}\}$
$(l=1,\ldots,n)$ and are independent of $j$ in
(\ref{eq:a-derivation}). One can easily obtain $f_k$ for the first
and last terms,
\begin{eqnarray}
f_1=(a(\lambda,z))^{(n)},\quad\quad f_n=((a(\lambda,z))')^{n}.
\end{eqnarray}
Thus, by means of the properties of determinants, the partition
function of the $gl(1|1)$ vertex model is simplified to the following
function:
\begin{eqnarray}
Z_N={(-\eta)^{\sum_{i=1}^{N-1}i}\;(b(\lambda,z))^N\over
 \prod_{i=2}^{N-1} i!}\,\prod_{i=2}^{N-1}i!\;
 \left[(a(\lambda,z))'\right]^{\sum_{i=1}^{N-1}i}
=(b(\lambda,z))^{N^2}.
\end{eqnarray}

\sect{$gl(2|1)$ vertex model with DW boundary condition}

\subsection{Description of the model}

Let $R\in End(V\otimes V)$ be the $R$-matrix associated with the
3-dimensional irreducible  $gl(2|1)$ module. Choosing the FFB grading
for $V$, i.e.  $[1]=[2]=1,[3]=0$, then the $R$-matrix reads
\begin{eqnarray}
 R_{12}(\lambda_1,\lambda_2)%\nonumber\\
% &=& \nonumber\\
          &=&\left(\begin{array}{ccccccccc}
 c_{12}&0&0 & 0&0&0 & 0&0&0\\
 0&a_{12}&0 & -b_{12}&0&0 & 0&0&0 \\
 0&0&a_{12} & 0&0&0  & b_{12}&0&0\\
 0&-b_{12}&0 & a_{12}&0&0  & 0&0&0\\
 0&0&0      & 0&c_{12}&0  & 0&0&0\\
 0&0&0      & 0&0&a_{12}  & 0&b_{12}&0\\
 0&0&b_{12} & 0&0&0       & a_{12}&0&0\\
 0&0&0      & 0&0&b_{12}  & 0&a_{12}&0\\
 0&0&0      & 0&0&0       & 0&0&1
 \end{array} \right), \label{de:R}
\end{eqnarray}
which satisfies the GYBE (\ref{eq:YBE}). Here $a_{12}, b_{12}$ and
$c_{12}$ are the same as those given in the previous section. The
basis vectors $|0\rangle,|1\rangle$ and $|2\rangle$ of the 3-d
$gl(2|1)$ representation are given by
\begin{eqnarray}
|0\rangle =\left(\begin{array}{c} 0\\0\\1\end{array}\right),
 \quad\quad
|1\rangle =\left(\begin{array}{c} 1\\0\\0\end{array}\right),
 \quad\quad
|2\rangle =\left(\begin{array}{c} 0\\1\\0\end{array}\right).
\end{eqnarray}

Now we consider the $gl(2|1)$ 15-vertex model on a $N\times N$
square lattice. Excluding the double occupancy, there are three
possible electronic states, i.e. up spin $\up$, down spin $\down$
and vacuum $\phi$, at each site of the lattice. For this model,
the configuration of the vertex is decided by the electronic
states around it. Corresponding to the $gl(2|1)$ invariance, there
are altogether 15 possible configurations, which preserve the
fermion numbers and spins, with non-zero Boltzmann weights. They
are shown by using figures as follows.
$$
%%%%%%%%%%%%%%%%%%%%%%%%%%%%%%%%%%%%%%%%%%%% line 1
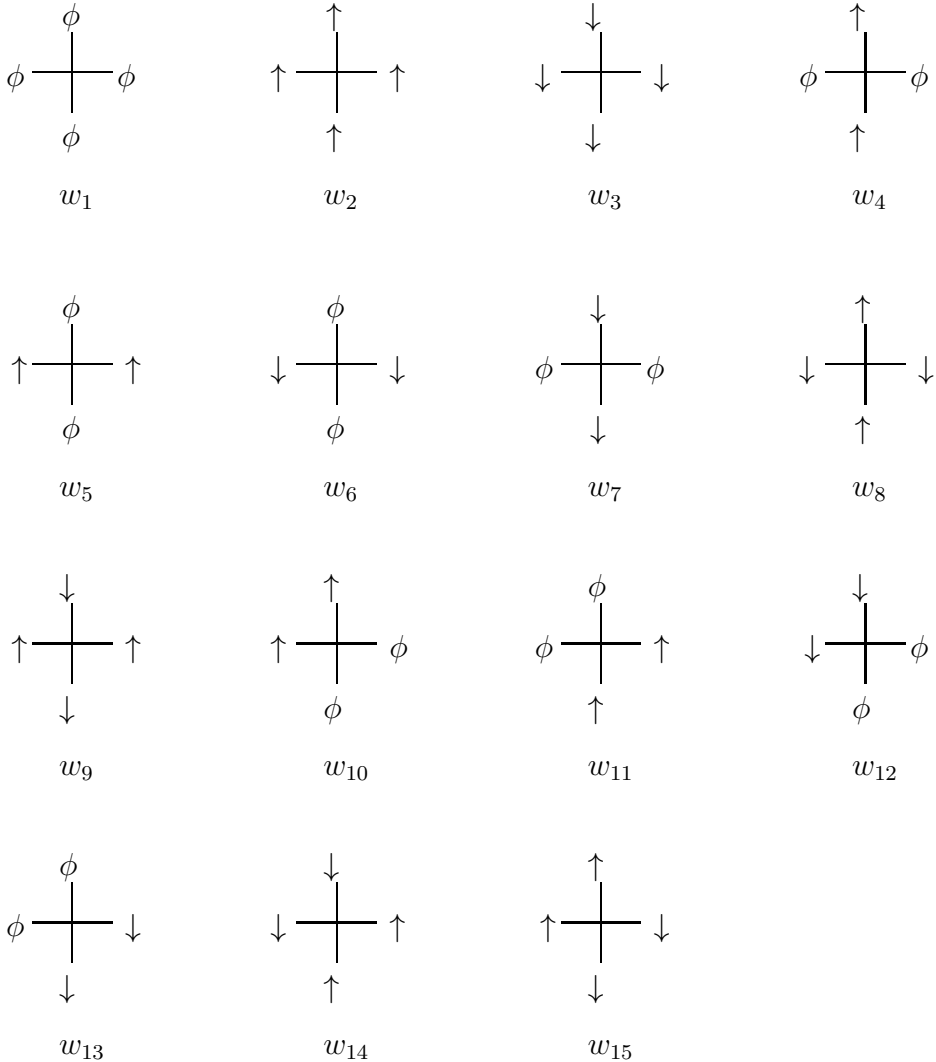
\begin{picture}(50,50)(0,0)
\put(-135,5){\line(-1,0){30}} \put(-150,20){\line(0,-1){30}}
\put(-154,-23){$\phi$}\put(-154,23){$\phi$}
\put(-175,0){$\phi$}\put(-133,0){$\phi$} \put(-155,-45){$w_1$}
\put(-35,5){\line(-1,0){30}} \put(-50,20){\line(0,-1){30}}
\put(-54,-23){$\up$}\put(-54,23){$\up$}
\put(-75,0){$\up$}\put(-30,0){$\up$} \put(-55,-45){$w_2$}
\put(65,5){\line(-1,0){30}} \put(50,20){\line(0,-1){30}}
\put(44,-23){$\down$}\put(44,23){$\down$}
\put(25,0){$\down$}\put(70,0){$\down$} \put(45,-45){$w_3$}
\put(165,5){\line(-1,0){30}} \put(150,20){\line(0,-1){30}}
\put(144,-23){$\up$}\put(144,23){$\up$}
\put(125,0){$\phi$}\put(167,0){$\phi$} \put(145,-45){$w_4$}
\end{picture}
$$\\[3mm]
%%%%%%%%%%%%%%%%%%%%%%%%%%%%%%%%%%%%%%%%%%%%%%%%%%%% line 2
$$
\begin{picture}(50,50)(0,0)
\put(-135,5){\line(-1,0){30}} \put(-150,20){\line(0,-1){30}}
\put(-154,-23){$\phi$}\put(-154,23){$\phi$}
\put(-173,0){$\up$}\put(-130,0){$\up$} \put(-155,-45){$w_5$}
\put(-35,5){\line(-1,0){30}} \put(-50,20){\line(0,-1){30}}
\put(-54,-23){$\phi$}\put(-54,23){$\phi$}
\put(-75,0){$\down$}\put(-30,0){$\down$} \put(-55,-45){$w_6$}
\put(65,5){\line(-1,0){30}} \put(50,20){\line(0,-1){30}}
\put(46,-23){$\down$}\put(46,23){$\down$}
\put(25,0){$\phi$}\put(67,0){$\phi$} \put(45,-45){$w_7$}
\put(165,5){\line(-1,0){30}} \put(150,20){\line(0,-1){30}}
\put(146,-23){$\up$}\put(146,23){$\up$}
\put(125,0){$\down$}\put(170,0){$\down$} \put(145,-45){$w_8$}
\end{picture}
$$\\[3mm]
%%%%%%%%%%%%%%%%%%%%%%%%%%%%%%%%%%%%%%%%%%%%%%%%%%%%
$$
\begin{picture}(50,50)(0,0)
\put(-135,5){\line(-1,0){30}} \put(-150,20){\line(0,-1){30}}
\put(-155,-23){$\down$}\put(-155,23){$\down$}
\put(-173,0){$\up$}\put(-130,0){$\up$} \put(-155,-45){$w_9$}
\put(-35,5){\line(-1,0){30}} \put(-50,20){\line(0,-1){30}}
\put(-55,-23){$\phi$}\put(-55,23){$\up$}
\put(-75,0){$\up$}\put(-30,0){$\phi$} \put(-55,-45){$w_{10}$}
\put(65,5){\line(-1,0){30}} \put(50,20){\line(0,-1){30}}
\put(45,-23){$\up$}\put(45,23){$\phi$}
\put(25,0){$\phi$}\put(70,0){$\up$} \put(45,-45){$w_{11}$}
\put(165,5){\line(-1,0){30}} \put(150,20){\line(0,-1){30}}
\put(145,-23){$\phi$}\put(145,23){$\down$}
\put(127,0){$\down$}\put(167,0){$\phi$} \put(145,-45){$w_{12}$}
\end{picture}
$$\\[3mm]
%%%%%%%%%%%%%%%%%%%%%%%%%%%%%%%%%%%%%%%%%%%%%%%%%%%%
$$
\begin{picture}(50,50)(0,0)
\put(-135,5){\line(-1,0){30}} \put(-150,20){\line(0,-1){30}}
\put(-155,-23){$\down$}\put(-155,23){$\phi$}
\put(-175,0){$\phi$}\put(-130,0){$\down$} \put(-155,-45){$w_{13}$}
\put(-35,5){\line(-1,0){30}} \put(-50,20){\line(0,-1){30}}
\put(-55,-23){$\up$}\put(-55,23){$\down$}
\put(-75,0){$\down$}\put(-30,0){$\up$} \put(-55,-45){$w_{14}$}
\put(65,5){\line(-1,0){30}} \put(50,20){\line(0,-1){30}}
\put(45,-23){$\down$}\put(45,23){$\up$}
\put(27,0){$\up$}\put(70,0){$\down$} \put(45,-45){$w_{15}$}
\end{picture}
$$\\[3mm]
$$
\begin{picture}(50,50)(0,0)
\put(-175,30){\small{\bf Figure 3.} Vertex configurations of the
$gl(2|1)$ vertex model and their Boltzmann weights.}
\end{picture}
$$
\vskip3mm
$$
\begin{picture}(50,50)(0,0)
\put(20,30){\line(-1,0){140}} \put(20,10){\line(-1,0){140}}
\put(20,-10){\line(-1,0){140}} \put(20,-30){\line(-1,0){140}}
\put(20,-50){\line(-1,0){140}} \put(20,-70){\line(-1,0){140}}
 \put(0,50){\line(0,-1){140}}\put(-20,50){\line(0,-1){140}}
 \put(-40,50){\line(0,-1){140}}\put(-60,50){\line(0,-1){140}}
 \put(-80,50){\line(0,-1){140}}\put(-100,50){\line(0,-1){140}}
 \put(-135,27){$\phi$} \put(-132,7){$\vdots$}
 \put(-135,-13){$\phi$} \put(-135,-33){$\phi$}
 \put(-132,-53){$\vdots$} \put(-135,-73){$\phi$}
 \put(27,27){$\down$}\put(29,7){$\vdots$}
 \put(27,-13){$\down$}\put(27,-33){$\up$}
  \put(29,-53){$\vdots$}\put(27,-73){$\up$}
 \put(-5,57){$\phi$}\put(-25,57){$\cdots$}
 \put(-45,57){$\phi$}\put(-65,57){$\phi$}
 \put(-85,57){$\cdots$}\put(-105,57){$\phi$}
 \put(-5,-105){$\up$} \put(-25,-105){$\cdots $}
 \put(-45,-105){$\up$} \put(-65,-105){$\down$}
 \put(-85,-105){$\cdots$} \put(-105,-105){$\down$}
\put(-110,42){{\small$z_1$}}\put(-93,42){{\small$\cdots$}}
\put(-75,42){{\small$z_P$}}\put(-58,42){{\scriptsize$z_{P+1}$}}
\put(-35,42){{\small$\cdots$}}\put(-12,42){{\small$z_N$}}
\put(-115,33){{\small$\l_1$}}\put(-113,13){{\small$\vdots$}}
\put(-113,-5){{\small$\l_P$}}\put(-119,-27){{\scriptsize$\l_{P+1}$}}
\put(-113,-47){{\small$\vdots$}}\put(-113,-67){{\small$\l_{N}$}}
%
%\put(-150,20){{$\left\{\begin{array}{c}\mbox{}\\
%  \mbox{}\end{array}\right.$}}
\end{picture}
$$ \vskip32mm
$$
\begin{picture}(50,50)(0,0)
\put(-190,30){\small{\bf Figure 4.}
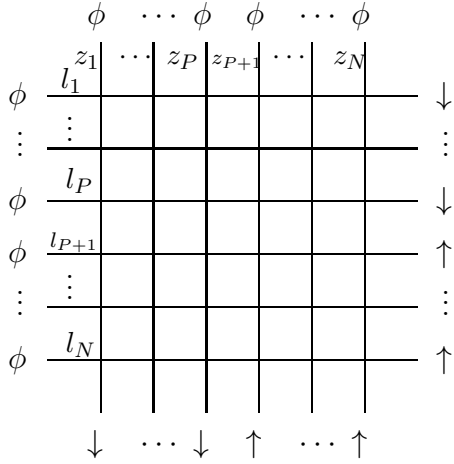
DW boundary condition for the $gl(2|1)$ vertex model.}
\end{picture}
$$
The Boltzmann weights associated with the vertices are given by
the elements of the $R$-matrix (\ref{de:R})
\begin{eqnarray}
&&w_1=1,\quad w_2=w_3=c(\lambda,z), \quad
w_4=\ldots=w_9=a(\lambda,z),\nonumber\\ &&
w_{10}=\ldots=w_{13}=b(\lambda,z),\quad
 w_{14}=w_{15}=-b(\lambda,z).
\label{de:w-tj}
\end{eqnarray}
%Thus, as before any horizontal or vertical line will form a
%monodromy matrix (\ref{de:T-tj}) and the Hamiltonian of the
%corresponding 1-dimensional system will be the supersymmetric $t$-$J$ model.

The $gl(2|1)$ supersymmetric vertex model with DW  boundary
condition is defined as follows. At the left and top ends, all
electrons are in vacuum states while at the right and bottom
ends, there are $P$ down spin states (corresponding to the
$1^{\mbox{st}}$-$P^{\mbox{th}}$ lines) and $N-P$ up spin states
(corresponding to the $(P+1)^{\mbox{th}}$ -$N^{\mbox{th}}$ lines).
The boundary condition is shown in figure 4. The DW partition
function of the $gl(2|1)$ vertex model is given by
\begin{eqnarray}
Z_N&=&\sum\prod_{i=1}^{15} w_i^{n_i} \nonumber\\
 &=&\prod_{i=0}^{P-1} \up_{(N-i)}
    \prod_{j=0}^{P-1}\down_{(P-j)}
    \prod_{i=0}^{N-1}\phi_{(N-i)}
    \prod_{k=1}^N\prod_{l=1}^N R_{kl}(\lambda_k,z_l)
    \prod_{j=1}^{P} \down_{(j)}
    \prod_{j=P+1}^{N}\up_{(j)}
    \prod_{i=1}^{N}\phi_{(i)}\nonumber\\
 &=&\prod_{i=0}^{P-1} \up_{(N-i)}
    \prod_{j=0}^{P-1}\down_{(P-j)}
    \prod_{i=0}^{N-1}\phi_{(N-i)}
    \prod_{k=1}^N T_k(\lambda_k)
    \prod_{j=1}^{P}\down_{(j)}
    \prod_{j=P+1}^{N}\up_{(j)}
    \prod_{i=1}^{N}\phi_{(i)}\nonumber\\
&=&\prod_{i=0}^{P-1} \up_{(N-i)}
    \prod_{j=0}^{P-1}\down_{(P-j)}
%    \prod_{i=0}^{N-1}\langle 0|_{(N-i)}
%    \prod_{k=1}^N\prod_{l=1}^N R_{kl}(\l_k,z_l)
%   \prod_{j=1}^{P}|\down\rangle_{(j)}
%  \prod_{j=P+1}^{N}|\up\rangle_{(j)}
    C_2(\lambda_1)\ldots C_2(\lambda_P)C_1(\lambda_{P+1})\ldots C_1(\lambda_N)
    \prod_{i=1}^{N}\phi_{(i)},\nonumber\\ \label{eq:Z-tj}
\end{eqnarray}
where $n_i$ are the number of configurations with the weights
$w_i$, and $T_k(\lambda_k)$ is the monodromy
 matrix  along the horizontal lines and is defined by
\begin{eqnarray}
T_k(\lambda_k)&=&R_{k,N}(\lambda_k,z_N)R_{k,N-1}(\lambda_k,z_{N-1})\ldots
R_{k,1}(\lambda_k,z_1)\nonumber\\ &\equiv&
 \left(\begin{array}{ccc}
 A_{11}(\lambda)k)&A_{12}(\lambda_k)& B_1(\lambda_k)\\
 A_{21}(\lambda_k)&A_{22}(\lambda_k)&B_{2}(\lambda_k)\\
 C_1(\lambda_k)&C_2(\lambda_k)&D(\lambda_k)
 \end{array}\right)_{(k)}. \label{de:T-tj}
\end{eqnarray}
Similar to the $gl(1|1)$ vertex model case, the partition function
(\ref{eq:Z-tj}) can be computed by using the approach of Drinfeld twists,
as can be seen from the next subsection.

\subsection{Exact solution of the partition function}

We now compute the partition function (\ref{eq:Z-tj})  using
the Drinfeld twist method. The $F$-matrix for this case is still defined by
(\ref{de:F}), except that now $\a=1,2,3$.
The inverse of the $F$-matrix is given by
\begin{equation}
F^{-1}_{1\ldots N}=F^*_{1\ldots N}\prod_{i<j}\Delta_{ij}^{-1}
\label{eq:F-inverse-tj}
\end{equation}
with
\begin{eqnarray}
\Delta_{ij}=\mbox{diag}\left(4a_{ij}a_{ji},a_{ji},a_{ji},a_{ij},4a_{ij}a_{ji},
a_{ji},a_{ij},a_{ij},1\right).
\end{eqnarray}

Working in the $F$-basis, the lower entries $C_1$ and $C_2$ of the
monodromy matrix $(\ref{de:T-tj})$ are simplified to symmetry
forms, that is they can be written as \cite{zsy05}
\begin{eqnarray}
\tilde C_2(\lambda) & =&\sum_{i=1}^Nb(\lambda,z_i)E^{2,3}_{(i)}
 \otimes_{j\ne i}\mbox{diag}\left(a(\lambda,z_j),2a(\lambda,z_j),1\right)_{(j)},
 \nonumber\\ \label{eq:C2-tilde}\\
\tilde C_1(\lambda)
 &=&\sum_{i=1}^N b(\lambda,z_i)
 E^{1,3}_{(i)}
 \otimes_{j\ne i}\mbox{diag}\left(2a(\lambda,z_j),
 a(\lambda,z_j)(a(z_i,z_j))^{-1},1\right)_{(j)}\nonumber\\
 & &+\sum_{i\ne j=1}^N
  {a(\lambda,z_i)b(\lambda,z_j)b(z_i,z_j)
   \over a(z_i,z_j)}
   E^{1,2}_{(i)} E^{2,3}_{(j)}
  \nonumber\\&&\quad
  \otimes_{k\ne i,j}\left(2a(\lambda,z_k),a(\lambda,z_k)a^{-1}(z_i,z_k),1
 \right)_{(k)}. \label{eq:C1-tilde}
\end{eqnarray}
Here $E^{\alpha\beta}_{(l)}$ ($\alpha, \beta = 1,2,3; l=1,2,\cdots, N$)
are the generators of $gl(2|1)$ on the $l$th vertical line.

In the following we will identify $|0\rangle, |1\rangle$ and
$|2\rangle$ with the vacuum $\phi$, spin-up $\uparrow$ and
spin-down $\downarrow$ states, respectively.  Note that under the
action of $F$ the state $|0\rangle_{(1)}\otimes\cdots\otimes
|0\rangle_{(N)}$ is invariant, that is,
\begin{eqnarray}
&& F_{1\ldots N}|0\rangle_{(1)}\otimes\cdots\otimes
  |0\rangle_{(N)}=|0\rangle_{(1)}\otimes\cdots\otimes |0\rangle_{(N)}.
  \label{eq:F-tj}
\end{eqnarray}
Thus substituting (\ref{eq:C2-tilde})-(\ref{eq:F-tj}) into the partition
function (\ref{eq:Z-tj}), we obtain
\begin{eqnarray}
Z_N&=&\prod_{i=0}^{P-1}\langle 1|_{(N-i)}
    \prod_{j=0}^{P-1}\langle 2|_{(P-j)}
    C_2(\lambda_1)\ldots C_2(\lambda_P)C_1(\lambda_{P+1})\ldots C_1(\lambda_N)
    \prod_{i=1}^{N}|0\rangle_{(i)}\nonumber\\
&=&\prod_{i=0}^{P-1}\langle 1|_{(N-i)}
    \prod_{j=0}^{P-1}\langle 2|_{(P-j)} F^{-1}_{1\ldots N}
    \tilde C_2(\lambda_1)\ldots \tilde C_2(\lambda_P)
    \tilde C_1(\lambda_{P+1})\ldots \tilde C_1(\lambda_N)
    \prod_{i=1}^{N}|0\rangle_{(i)}\nonumber\\
&=&\sum_{i_1<\ldots<i_{P}}\sum_{i_{P+1}<\ldots<i_{N}}
 \prod_{l=0}^{P-1}\langle1|_{(N-l)}
 \prod_{j=0}^{P-1}\langle 2|_{(P-j)} F^{-1}_{1\ldots N}
  {\prod_{j=1}^{P}}E^{23}_{(i_j)}
 {\prod_{j=P+1}^{N}}E^{13}_{(i_j)}\prod_{l=0}^N|0\rangle_{(l)}
\nonumber\\ &&\times
 2^{{P(P+1)\over 2}+{(N-P)(N-P+1)\over2}}
 \prod_{k=1}^{P}\prod_{l=P+1}^N a(\lambda_{k},z_{i_l})
\nonumber\\ &&\times
 \mbox{det}{\cal B}_P(\{\lambda_1,\ldots,\lambda_P\},
    \{z_{i_1}\ldots,z_{i_P}\})\nonumber\\
&&\times \mbox{det}{\cal
B}_{N-P}(\{\lambda_{P+1},\ldots,\lambda_N\},
  \{z_{i_{P+1}}\ldots,z_{i_N}\})\nonumber\\
&\equiv&\sum_{i_1<\ldots<i_{P}}\sum_{i_{P+1}<\ldots<i_{N}}
 \prod_{l>k=1}^{P}(2a(z_{i_k},z_{i_l}))^{-1}
    \prod_{l>k=P+1}^{N}(2a(z_{i_k},z_{i_l}))^{-1}
    G(\{\underline{z}\})
\nonumber\\ &&\times
 2^{{P(P+1)\over 2}+{(N-P)(N-P+1)\over2}}
 \prod_{k=1}^{P}\prod_{l=P+1}^N a(\lambda_{k},z_{i_l})
\nonumber\\ &&\times
 \mbox{det}{\cal B}_P(\{\lambda_1,\ldots,\lambda_P\},
    \{z_{i_1}\ldots,z_{i_P}\})\nonumber\\
&&\times \mbox{det}{\cal
B}_{N-P}(\{\lambda_{P+1},\ldots,\lambda_N\},
  \{z_{i_{P+1}}\ldots,z_{i_N}\}) , \label{Z-tj}
\end{eqnarray}
where $\{i_1,\ldots,i_P\}\cap\{i_{P+1},\ldots,i_n\}=\varnothing$
and ${\cal B}_M(\{\underline{\lambda}\},\{\underline{z}\})$ is a
$M\times M$ matrix with elements
\begin{equation}
({\cal
B}_M(\{\underline{\lambda}\},\{\underline{z}\}))_{\alpha\beta}
 =b(\lambda_\alpha,z_\beta)
  \prod_{\gamma=1}^{\alpha-1}a(\lambda_\gamma,z_\beta).
\end{equation}
In (\ref{Z-tj}), the function $G(\{\underline z\})$ is defined by
\begin{eqnarray}
G(\{\underline z\})&\equiv& \prod_{l>k=1}^{P}(2a(z_{i_k},z_{i_l}))
    \prod_{l>k=P+1}^{N}(2a(z_{i_k},z_{i_l}))
    \langle 1|_{(N)}\ldots\langle 1|_{(P+1)}
 \langle 2|_{{(P)}}\ldots\langle 2|_{(1)}\,\,
  \nonumber\\
 &&\times
 F^{-1}_{1\ldots N}
  E^{23}_{(i_1)}\ldots E^{23}_{i_P} E^{13}_{(i_{P+1})\ldots}E^{13}_{(i_N)}\,\,
 |0\rangle_{(1)}\ldots|0\rangle_{(N)}
\end{eqnarray}
Using the matrix $F^{-1}_{1\ldots N}$ defined by
(\ref{eq:F-inverse})-(\ref{cond:F*}), one may simplify
$G(\{\underline z\})$ as follows.
\begin{eqnarray}
G(\{\underline z\})&=&\prod_{l>k=1}^{P}(2a(z_{i_k},z_{i_l}))
    \prod_{l>k=P+1}^{N}(2a(z_{i_k},z_{i_l})) \nonumber\\
 &&\times
 \langle0|_{(N)}\ldots
 \langle0|_{(1)}E^{31}_{(N)}\ldots E^{31}_{(P+1)}
   E^{32}_{(P)}\ldots E^{32}_{(1)}\,\,
 \sum_{\sigma\in {\cal S}_2} S(c,\sigma,\alpha_\sigma)
 \nonumber\\
 && \times R^{\sigma^{-1}}_{\sigma(1\ldots N)}P_{(\sigma(1))}^{\alpha_{\sigma(1)}}
  \ldots P_{(\sigma(N))}^{\alpha_{\sigma(N)}} \prod_{i<j}
  \Delta^{-1}_{ij}\,\,
  E^{23}_{(i_1)}\ldots E^{23}_{i_P} E^{13}_{(i_{P+1})\ldots}E^{13}_{(i_N)}\,\,
 |0\rangle_{(1)}\ldots|0\rangle_{(N)}
 \nonumber\\
 &=&
 %\lim_{z_{1},\ldots,z_{N}\rightarrow z}
 %\sum_{i_1<\ldots<i_{P}}\sum_{i_{P+1}<\ldots<i_{N}}
 \prod_{k=1}^P\prod_{l=P+1}^N a^{-1}(z_{i_k},z_{i_l})% \nonumber\\
% &&\times
\prod_{k>l=1}^{P}(2a(z_{i_k},z_{i_l}))^{-1}
    \prod_{k>l=P+1}^{N}(2a(z_{i_k},z_{i_l}))^{-1} \nonumber\\
 &&\times
 \langle0|_{(N)}\ldots
 \langle0|_{(1)}E^{31}_{(N)}\ldots E^{31}_{(P+1)}
   E^{32}_{(P)}\ldots E^{32}_{(1)}\,\,
 \sum_{\sigma\in {\cal S}_N} S(c,\sigma,\alpha_\sigma)
 \nonumber\\
 && \times R^{\sigma^{-1}}_{\sigma(1\ldots N)}
  E^{23}_{(i_1)}\ldots E^{23}_{i_P} E^{13}_{(i_{P+1})\ldots}E^{13}_{(i_N)}\,\,
 |0\rangle_{(1)}\ldots|0\rangle_{(N)}\nonumber\\
&=& \prod_{k=1}^P\prod_{l=P+1}^N a^{-1}(z_{i_k},z_{i_l})% \nonumber\\
% &&\times
\prod_{k>l=1}^{P}(2a(z_{i_k},z_{i_l}))^{-1}
    \prod_{k>l=P+1}^{N}(2a(z_{i_k},z_{i_l}))^{-1} \nonumber\\
 &&\times  \sum_{\sigma\in {\cal
 S}_N}(-1)^{\mbox{sign}(\sigma(i_1,\ldots,i_N)=(1,\ldots,N))}
 S(c,\sigma,\alpha_\sigma)
 \left(R^{\sigma^{-1}}_{\sigma(1\ldots N)}
 \right)_{2\ldots 21\ldots 1}^{\a_{1}\ldots \a_{P}
 \a_{P+1}\ldots\a_N},
  \label{eq:G-proof}
\end{eqnarray}
where $\a=1,2$, the subscribes of $\a$ are indices of space,
$\mbox{sign}(\sigma)=1$ if $\s$ is odd and $\mbox{sign}(\sigma)=0$
if $\s$ is even.

The procedure of computing the homogenous limit is similar to that
for the $gl(1|1)$ vertex model. Here we only give the results. In
the homogenous limit, i.e. when $\lambda_1=\ldots=\lambda_N\equiv
\lambda$ and $z_1=\ldots=z_N\equiv z $,  the partition function
(\ref{Z-tj}) becomes
\begin{eqnarray}
Z_N&=&(a(\lambda,z))^{P(N-P)}(b(\lambda,z))^{P^2}(b(\lambda,z))^{(N-P)^2}
\nonumber\\&&\times
 \lim_{z_{1},\ldots,z_{N}\rightarrow z}
 \sum_{i_1<\ldots<i_{P}}\sum_{i_{P+1}<\ldots<i_{N}}G(\{\underline{z}\})
 \label{eq:G1}
\end{eqnarray}

By using the decomposition law (\ref{eq:R-RR}), the $R$-matrix
$R^\s$ in (\ref{eq:G-proof}) can be decomposed to elementary
$R$-matrices. In homogenous limit, the elements of the elementary
$R$-matrix will tend to
\begin{eqnarray}
R(z,z)^{12}_{21}=R(z,z)^{21}_{12}= -1, \quad
R(z,z)^{11}_{11}=R(z,z)^{22}_{22}= -1, \quad
R(z,z)^{12}_{12}=R(z,z)^{21}_{21}= 0. \quad
\end{eqnarray}
Therefore for the last factor in (\ref{eq:G1}) involving
$G(\{\underline{z}\})$,
%%%%%%%%%%%%%%%%%%%%%%%%%%%%%%%%%%%%%%%%%%%%%%%%%%%%%%%%%%%%%%%%%%
%equals to the combinatorial number $C^P_N$. For example, when
%$N=3$ and $P=1$,
%%%%%%%%%%%%%%%%%%%%%%%%%%%%%%%%%%%%%%%%%%%%%%%%%%%%%%%%%%%%%%%%%%
one may easily obtain
\begin{eqnarray}
 && \lim_{z_{1},\ldots,z_{N}\rightarrow z}
 \sum_{i_1<\ldots<i_{P}}\sum_{i_{P+1}<\ldots<i_{N}}
   G(\{\underline{z}\})\nonumber\\
&=&\lim_{z_{1},\ldots,z_{N}\rightarrow z}
 \sum_{i_1<\ldots<i_{P}}\sum_{i_{P+1}<\ldots<i_{N}}
 \prod_{k=1}^P\prod_{l=P+1}^N a^{-1}(z_{i_k},z_{i_l}) \nonumber\\
&=&C_N^P,
\end{eqnarray}
where $C_N^P$ is the combinatorial number.

In summary, in the homogeneous limit the DW partition function $Z_N$ is
\begin{eqnarray}
Z_N=C^P_N\,(a(\lambda,z))^{P(N-P)}\,(b(\lambda,z))^{P^2+(N-P)^2}.
\end{eqnarray}

It is easily seen that when $P=0$ or $P=N$, the DW partition
function of the $gl(2|1)$ vertex model reduces to that of the
$gl(1|1)$ vertex model, confirming a statement made in \cite{Foda96}
on DW partition functions of vertex models based on $A_n$-type
algebras. However, when $P\neq 0, N$, one obtains new partition
functions of the true $gl(2|1)$ supersymmetric vertex model with
DW boundary conditions.

\sect{Conclusion and discussion}

In this paper, we have proposed the $gl(1|1)$ and $gl(2|1)$
supersymmetric vertex models with the so-called DW boundary
conditions. The DW partition functions of the models have been
computed by means of the approach of the Drinfeld twists. We have
found that in the homogenous limit, the partition functions
degenerate to simple functions. For the 2-d square $gl(2|1)$
lattice model, we note here that the definition of the DW boundary
condition is not unique. For the other cases, by using the same
procedure, one may find their partition functions are similar with
those in this paper.

We have demonstrated, by working out the DW $gl(1|1)$ and
$gl(2|1)$ supersymmetric vertex models as examples, that by means
of the Drinfeld twist method, one can actually derive directly,
rather than conjecture a formula and then verify it as usually
done, the determinant representations of DW partition functions.

It is widely known that determinant representations of partition
functions are closely related to some pure mathematical problems,
such as algebraic combinations and tilings of the Aztec diamond
\cite{Korepin00,Zinn-Justin00}. In our further work, we will study
the mathematical problems arising from the present models. The
results in this paper will also be useful for simplifying the
correlation functions of the supersymmetric $t$-$J$ model obtained
in {\cite{zsy0511}} and for further studying physical properties
of the model.

\vskip.3in \noindent {\bf Acknowledgements:}  SYZ was supported by
the UQ Postdoctoral Research Fellowship. SYZ would like to thank
W.-L. Yang and K.-J. Shi for helpful discussions. YZZ was
supported by Australian Research Council and also by
Max-Planck-Institut f\"ur Mathematik  and Alexander von
Humboldt-Stiftung. YZZ would like to thank the
Max-Planck-Institute f\"ur Mathematik, where part of this work was
done, and Physikalisches Institut der Universit\"at Bonn,
especially G\"unter von Gehlen, for warm hospitality.

\end{document}